\documentclass[12pt,a4paper,final]{iopart}

\usepackage{iopams} 
\usepackage{svg}
\expandafter\let\csname equation*\endcsname\relax
\expandafter\let\csname endequation*\endcsname\relax
\usepackage{amsmath}
\usepackage{graphicx}
\usepackage[breaklinks=true,colorlinks=true,linkcolor=blue,urlcolor=blue,citecolor=blue]{hyperref}
\usepackage{fancyhdr}
\usepackage{dcolumn}%
\usepackage{array}
\usepackage{multirow}
\usepackage{overpic}
\usepackage{subfig}
\usepackage{cite}
\usepackage{bm}%
\usepackage{tabularx}
\usepackage{booktabs}

\usepackage{float}
\usepackage{mathtools}
\usepackage{xfrac}
\expandafter\ifx\csname package@font\endcsname\relax\else
 \expandafter\expandafter
 \expandafter\usepackage
 \expandafter\expandafter
 \expandafter{\csname package@font\endcsname}%
\fi
\newcommand{\super}[1]{\ensuremath{^{\textrm{#1}}}}
\newcommand{\sub}[1]{\ensuremath{_{\textrm{#1}}}}
\newcommand{\tilda}[1]{\ensuremath{\sim{}}}

\colorlet{lightred}{red!35!white}
\colorlet{lightyellow}{yellow!35!white}
\colorlet{lightgreen}{green!35!white}

\DeclareSymbolFont{boldoperators}{OT1}{cmr}{bx}{n}
\SetSymbolFont{boldoperators}{bold}{OT1}{cmr}{bx}{n}

\begin{document}

\title[]{Magnetic properties of nickel electrodeposited on porous GaN substrates with infiltrated and laminated connectivity}

\author{Yana Grishchenko$^{1}$}
\author{Josh Dawson$^{2}$}
\author{Saptarsi Ghosh$^{2}$}
\author{Abhiram Gundimeda$^{2}$}
\author{Bogdan F. Spiridon$^{2}$}
\author{Nivedita L. Raveendran$^{1}$}
\author{Rachel A. Oliver$^{2}$}
\author{Sohini Kar-Narayan$^2$}
\author[cor1]{Yonatan Calahorra$^{1}$}
\address{$^1$Department of Materials Science and Engineering, Technion - IIT, Haifa, 3200003, Israel}
\address{$^2$Department of Materials Science and Metallurgy, University of Cambridge, CB3 0FS, Cambrdige, UK}
\eads{\mailto{calahorra@technion.ac.il}}

\begin{abstract}
We studied the magnetic properties of ferromagnetic-semiconductor composites based on nickel and porous-GaN, motivated by the effort to couple magnetic and semiconductor functionality. Nickel-infiltrated and nickel-coated (laminated thin-film) porous GaN structures were fabricated by electrodeposition, and their magnetic properties were subsequently examined collectively, by vibrating sample magnetometry and on the nanoscale, by magnetic force microscopy. We successfully demonstrated the ability to realize nickel infiltrated porous GaN, where the magnetic properties were dominated by the infiltrated material without a measurable surface contribution. We found that the structure and magnetization of electrodeposited porous-GaN/Ni composites depended on GaN degree of porosity and the amount of deposited nickel. The magnetization evolves from a nearly isotropic response in the infiltrated structures, to a shape-anisotropy controlled magnetic thin-film behaviour. Furthermore, both infiltrated and thin-film nickel electrodeposited on porous GaN were found to have low ($<$ 0.1\%) strain and corresponding low coercivity: $<$ 6.4 and $<$ 2.4 kA/m for infiltrated and thin-film, correspondingly. The most likely cause for the lowered strain is increased compliance of the porous GaN compared to bulk. These results encourage deeper investigation of magnetic nanostructure property tuning and of magnetic property coupling to GaN and similar materials.  
\end{abstract}

\maketitle
\section{Introduction}
Direct band-gap, non-centrosymmetric, semiconductors such as those of the III-V and II-VI families are multifunctional materials offering inherent coupling between mechanical, electrical, optical properties and semiconducting functionality\cite{Jenabook,Humphreys2008,Lagowski1972,kusaka1978electrical,Calahorrabook,Yang2010,Wang2014effects,keil2017piezotronic,Fromling2018piezotronic,Calahorra2019Highly}. In the past two decades a significant effort to introduce magnetic functionality to semiconductors has been ongoing, mainly motivated by spintronic and spinoptic applications \cite{Ohno1999electrical,Dietl2000zener}, as well as by a magnetoelectric driver: electrically controlled magnetism\cite{Ohno2000electric}. While the dominant route to achieve this property coupling is through dilute magnetic semiconductors\cite{Lin2017enhanced,Wadekar2019mn,Jiang2020single}, challenges in this technology (low Curie temperatures, solubility limitations) brought increasing interest in magnetic/semiconductor composites\cite{bonanni2007paramagnetic,Gerngross2014Dec,Lv2014feco,Navarro-Quezada2019Feb,Navarro2020out}. The importance of GaN as a semiconducting material makes it attractive for magnetic/semiconductor coupling. Indeed, various magnetic-GaN applications were demonstrated, including spintronic\cite{Zhang2021Electrical}, spin-optic\cite{Chen2014self} and magneto-electro-optical\cite{Wei2011optical,Chen2013optically}. The coupling between magnetic and semiconducting properties and the ability to control it are gaining increased attention, and various methods and geometries to realise the composites exist, including epitaxial growth and phase separation\cite{bonanni2007paramagnetic,Navarro-Quezada2019Feb,Navarro2020out}, laminate composites\cite{Wei2011optical,Gao2019evidence,Zhang2021Electrical}, nanostructure coupling\cite{Lv2014feco,Chen2014self} and templated deposition\cite{Gerngross2014Dec}.\\ \indent
Porous III-V semiconductors have unique mechanical and optical properties\cite{Huang2013mechanical,Zhang2015mesoporous,Zhu2017wafer} and they exhibit enhanced piezoelectricity\cite{Gerngross2011Apr,Waseem2018May,Calahorra2020enhanced}. The porous structure is attractive for examination of composite material combinations. Filling of porous GaN with an optoelectronic perovskite was recently demonstrated\cite{Lim2019Feb}, with the motivation of coupling two distinct optoelectronic materials. In a similar fashion to the presented work, filling a porous InP with magnetic material to form nanowires was reported\cite{Gerngross2011Apr,Gerngross2014Dec}, aiming at magnetoelectric applications. GaN offers better piezoelectric properties than InP and presents more opportunities for related property coupling. In the following, we describe a method for fabrication of porous GaN and nickel composites - by electrodeposition. We studied the magnetic properties of the composite structures and found that the magnetic response was dependent upon the characteristics of the porous layer and amount of deposited nickel, spanning nearly isotropic responses and transition to thin-film like behaviour. Overall, this study demonstrates the controlled synthesis of laminate as well as infiltrated Ni/GaN composites and the interplay between synthesis process and magnetic properties. The results drive the next steps in studying magnetic/semiconductor composites. \\ \indent
\section{Materials and Methods}
\subsection{GaN growth and electrochemical etching}
Si-doped GaN (n-type, sample doping ranged 5.5$\times$10\super{18} - 2.3$\times$10\super{19} cm\super{-3}; 1 $\mu$m thick) was grown on a low-dislocation GaN pseudo-substrate\cite{Datta2004growth} on sapphire by metal-organic vapor phase epitaxy (MOVPE) and electrochemically porosified. Electrochemical etching was carried out in a two-electrode configuration. The sample was immersed in oxalic acid (0.25M) with the nitride surface of the doped layer exposed and a DC potential was applied between the sample and a platinum counter electrode. In this process, the porosity degree follows the voltage, the depth follows time and the process as a whole depends upon doping (through the influence of sample conductivity and interfacial hole tunneling)\cite{Zhu2017wafer,Griffin2018porous}. The chosen doping level was $1.85\times$10\super{19} cm\super{-3} and the etching voltage was either 10 or 12 V; see further discussion in Supporting Information S1. In order to reuse the electrochemical sample configuration, the etch duration was controlled such that a part of the n-doped layer was left intact, with the purpose of acting as a working electrode for subsequent nickel electrodeposition (\Fref{fig:Scheme}).
\subsection{Nickel deposition}
\begin{figure}[h]
 \centering
 \includegraphics[scale=0.47]{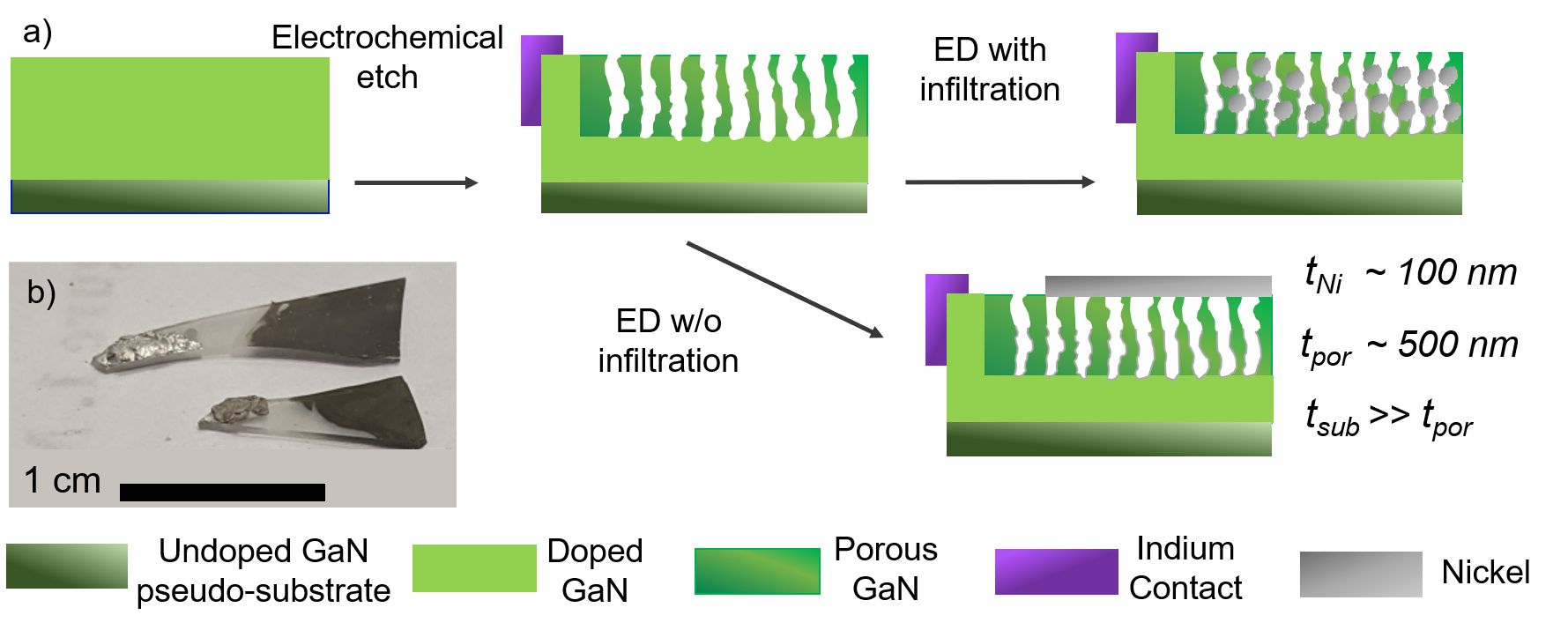}
 \caption{a) sample preparation schematic. A highly doped GaN layer was electrochemically etched to form porous structures. The same configuration was used to electrodeposit nickel within the pores. Depending on the porous layer conductivity and pore size, it was found that nickel either formed on top of the sample, or nucleated on the pore walls. The typical thickness scales of the process layers are shown; b) two electrodeposited samples.}\label{fig:Scheme}
\end{figure}

Nickel was electrodeposited on the etched substrates, resulting in Ni-infiltrated GaN, and Ni-coated GaN. Electrodeposition was carried out in a three electrode configuration\cite{Boughey2018coaxial}, using a counter Pt electrode, an Ag/AgCl reference electrode and the sample as a working electrode. A nickel sulfate (NiSO\sub{4}) and boric acid (H\sub{3}BO\sub{3}) solution, pH 4.5, was used. See complete details in Supporting information S2. For the selected working potential (-1 V), the deposition rate was linear with time (after initial current stabilisation). Following deposition, the samples were washed with DI water and dried under nitrogen flow.\\ \indent
\Fref{fig:Scheme} shows a schematic of the sample processing and its outcomes (as demonstrated by electron microscopy): electrochemical etching of conductive GaN followed by electrodeposition of nickel (using the same apparatus) and coated or infiltrated composites.\\
\subsection{Electron microscopy}
Scanning electron microscopy (SEM) characterization was carried out using a \textit{Zeiss Ultra Plus} operated at 2.5-4 kV. SEM and energy dispersive X-ray (EDX) spectroscopy measurements were also done on an \textit{FEI Nova NanoSEM} operated at 5 kV by imaging secondary electrons. The SEM is equipped with a silicon Drift Detector Energy Dispersive X-ray spectrometer.
\subsection{X-ray diffraction}
X-ray diffraction (XRD) measurements were carried out using a \textit{Rikagu SmartLab} 9 kW without sample rotation. The D/tex Ultra detector in 1D mode, Ge ((220)x2) monochromator, $\lambda$ = 1.541Å were used, and scanning was done at high resolution (0.01$^{\circ}$) steps, from 2$\theta$ = 43 - 53$^{\circ}$, where we expected to detect Ni signals\cite{Lantelme1998model,Boubatra2011morphology}. (200):(111) integrated peak ratios were calculated manually on the background subtracted data, accounting for cases where (200) peak was not detected by the fitting software.
\subsection{Magnetic characterization}
\subsubsection{Vibrating sample magnetometry.}
Temperature-dependent vibrating sample magnetometry (VSM) was carried out using a Physical Property Measurement System (PPMS) DynaCool (\textit{Quantum Design}) VSM option. The maximal DC H field applied was 10 kOe (1 T). The sample was mounted on a quartz holder for in-plane (IP) magnetization measurements and inside a plastic straw for out-of-plane (OOP) magnetization data. The diamagnetic/paramagnetic contributions of the holders were subtracted from the signal following linear fitting of the saturated regions. No demagnetization correction factor was used.
\subsubsection{Magnetic force microscopy.}
Magnetic force microscopy (MFM) was performed on a Bruker Icon Dimension atomic force microscope (AFM) in tapping-MFM and peak-force-MFM (PF-MFM) modes. Probes used were MESP-RC-V2 (Co/Cr coating) by Bruker. The probes were magnetized along the tip before the measurements. The samples were magnetized manually by carefully moving into and out-of the field of a commercial permanent neodymium magnet stack,  either out-of-plane or in-plane. All MFM measurements reported were taken using a second pass technique (after tapping/PF topography measurement) regardless of the sample magnetization. AFM image analysis was carried out on Bruker software Nanoscope Analysis 3. Topography and MFM images were flattened to reduce interline drifts and MFM images were cleaned from streaks - all by existing tools.  
\section{Results and discussion}
\subsection{Nickel deposition and resulting structures}
\begin{figure}[!h]
 \centering
 \includegraphics[scale=0.42]{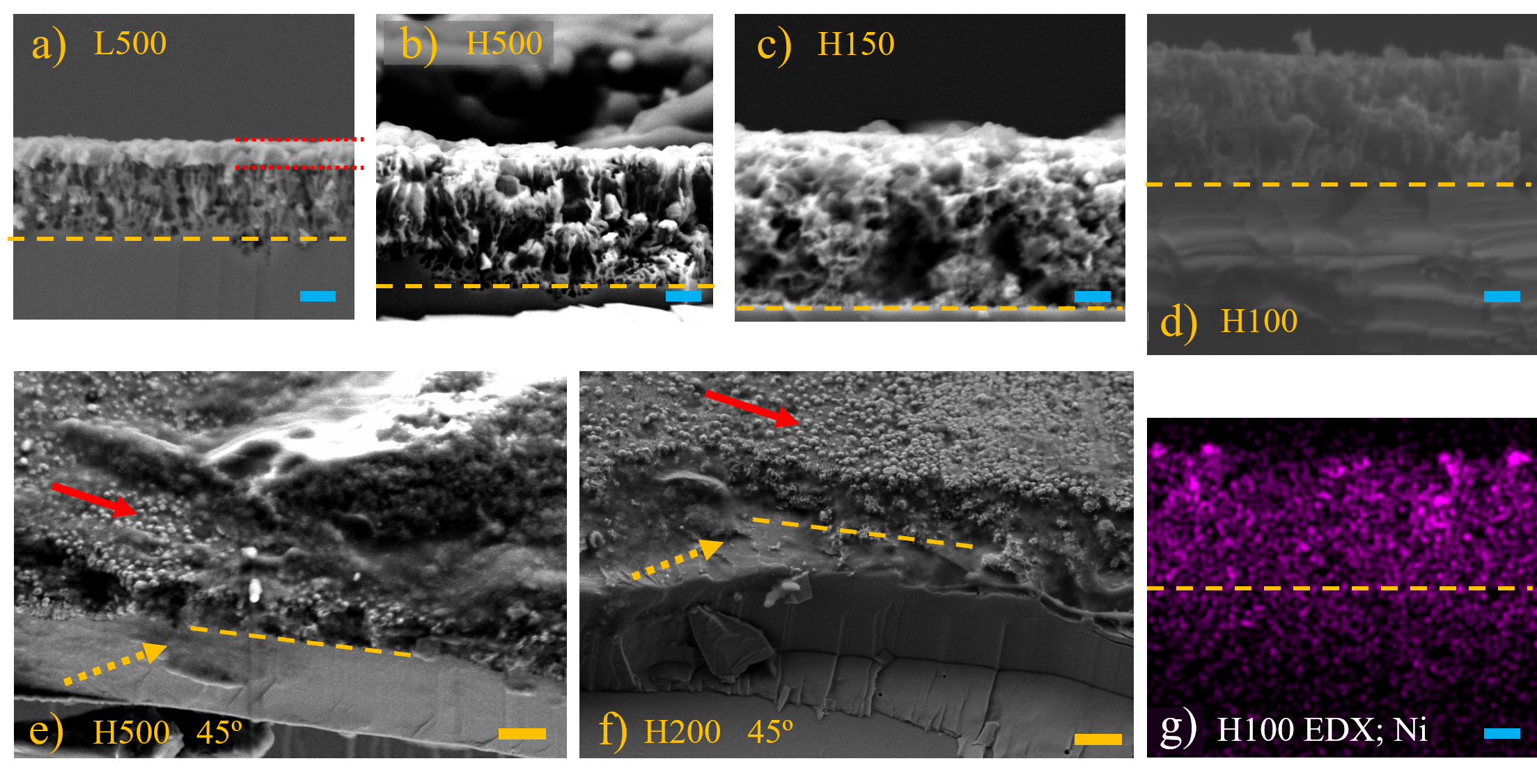}
 \caption{SEM images of nickel electrodeposition: a-d) Cross-sectional SEM of some of the samples considered in this study, samples marked on image. The dashed orange lines indicate the porous/non-porous boundary and the dotted red lines in [a] indicate the thickness of the nickel thin film atop L500; e) SEM image of H500 captured at 45$^{\circ}$; f) SEM image of H200 captured at 45$^{\circ}$; g) nickel EDX signal of H100, corresponding to [d]. Arrows point to the samples' top surfaces (solid red) and cross-sections (dashed orange). Scale bars are 200 nm, except [e,f] which are 800 nm.}\label{fig:SEM}
\end{figure}

\indent
Two degrees of porosity were examined in this work, controlled by the etching bias: 10 and 12 V, with an estimated (based on a previous study\cite{Calahorra2020enhanced}) porosity of 40-60\% and 60-80\%, and 30-50 nm and 50-100 nm sized pores, correspondingly. Electrodeposition on these porosified samples was carried out for a set period of time in the 100-500 s range. We designate the samples by their high/low etching voltage (12/10; H/L), and the deposition duration; \textit{e.g.} a 12 V etched sample (higher voltage and larger pore process) which underwent electrodeposition for 150 s is sample H150.\\ \indent
\Fref{fig:SEM} shows cross sectional or tilted SEM images of several samples and the nickel EDX signal of sample H100. An expected increased pore size of the 12 V samples was observed (\textit{e.g.\nolinebreak\ }\Fref{fig:SEM}b-d compared to \Fref{fig:SEM}a). In addition, the electrodeposited Ni formed strikingly different structures on larger/smaller pores: sample H500 displayed a combination of infiltrated nickel (\Fref{fig:SEM}b bright contrast) and top coverage. \Fref{fig:SEM}e shows a lower magnification 45$^{\circ}$ capture at the edge of H500 showing both the top surface (solid arrow) and cross section (dashed arrow), displaying a rough surface, with protrusions forming an uneven surface. The effect of deposition time is demonstrated through a further comparison to sample H200 which exhibited a similar general appearance, with significantly less surface growth and thickening of protrusions (\Fref{fig:SEM}f). When comparing H500 to its smaller-pore counterpart, sample L500 showed mostly a thin-film forming on top of the porous GaN, with only a few nickel nanoparticles forming inside the pores (\Fref{fig:SEM}a).\\ \indent
This finding can be attributed to a relation between pore size and porous layer conductivity: the epitaxial GaN layer was highly doped. A higher-bias etching (12  compared to 10 V) left behind less material, resulting in larger pores and higher resistivity, which overall encouraged infiltration. We conclude that the 10 V etching process left the surface sufficiently conductive to allow for electrodeposition directly on the surface - facilitating the deposition of a thin film.\\ \indent
On the other hand the 12 V process resulted in the formation of a Ni-infiltrated structure, evident by the absence of a thin-film on top of samples H150 and H100 (\Fref{fig:SEM}c,d). This indicates that the samples surfaces of 12 V samples was sufficiently non conductive to favor electrodeposition nucleation within the porous GaN, and that subsequent surface coating is dominated by nickel protrusion outside the surface. Furthermore, SEM-EDX measurements on sample H100 (\Fref{fig:SEM}g and Supporting Information Figure S2.3), revealed the presence of nickel throughout the porous layer, although no clear contrast between deposition and matrix was visible in the secondary electrons image itself. It is also possible that hydrogen bubbles formed (in a competing route to Ni\cite{Lantelme1998model}) during electrodeposition prevented the formation of bottom up growth in the porous, confined, space\cite{Boubatra2011morphology}. Both infiltrated and top-cover electrodeposition routes are depicted in \Fref{fig:Scheme}.\\ \indent

\begin{figure}[h]
 \centering
 \includegraphics[scale=0.47]{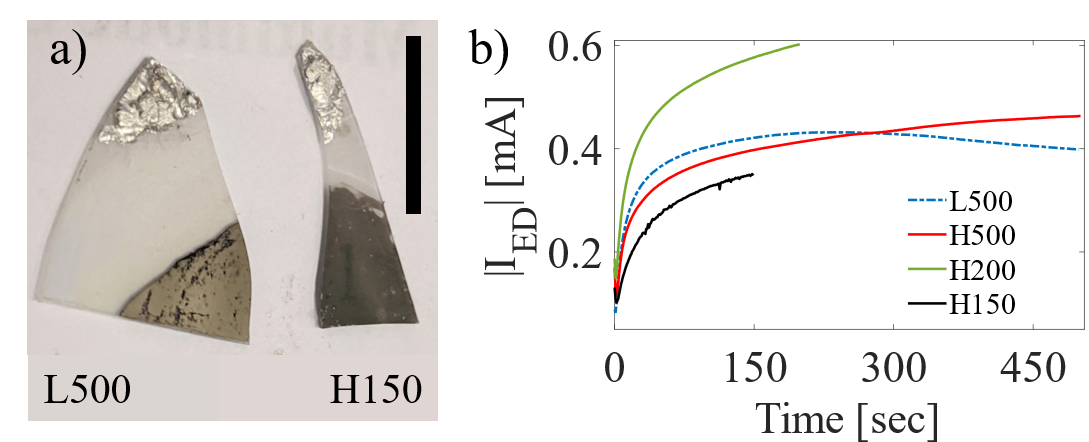}
 \caption{a) An optical image of some of the resulting structures. The indium contact and the border between deposited areas (immersed in the solution) and deposition-free areas are visible. Sample L500 exhibited a lustrous finish, corresponding to a mostly top deposition while sample H150 seems more opaque, corresponding to an infiltrated deposition. Scale bar is 1 cm; b) The currents recorded during electrodeposition. The difference between H500 and L500 is noticeable. The current was not normalized, reflecting different surface areas of the samples.}\label{fig:ED_currents}
\end{figure}

The visual appearance of the samples reflected their structure (\Fref{fig:ED_currents}a): 12 V etched samples with infiltrated deposition (\textit{e.g.} H150), displayed a matte finish, while the L500 sample, with mostly surface deposition, displayed a lustrous finish. When considering the current during electrodeposition (\Fref{fig:ED_currents}b), it is interesting to compare the two 500 s samples: L500 showed a current maximum around 150 sec, while the current did not reach a maximum in H500. Supporting Information Figure S2.1 shows control depositions on ITO (indium tin oxide) substrates, which lasted longer than 500 s and showed the current mostly saturated after 500 sec, and did not drop. We suggest that the porous substrate plays an interesting role in the evolution of the electrodeposition front-line. When considering the conductive, non-porous, ITO substrates, the deposition nucleates at preferential points, then expands until a stable area and cover is achieved and maintained - resulting in a saturated current\cite{Lantelme1998model}. The H500 sample exhibited similar profiles, however did not reach saturation in 500 sec. This is possibly due to the larger surface area, leading to larger (volume) density of nucleation points for electrodeposition, combined with 3D growth of the nuclei. The L500 sample displayed a different trend and reached a relatively fast maximum. A possible explanation is an initial coexistence of surface and pore nucleation and deposition, where as the pores near the surface were covered, deposition within the pores stopped, reducing the effective area for electrodeposition and correspondingly, the current.\\ \indent  

\begin{figure}[!h]
 \centering
 \includegraphics[scale=1.1]{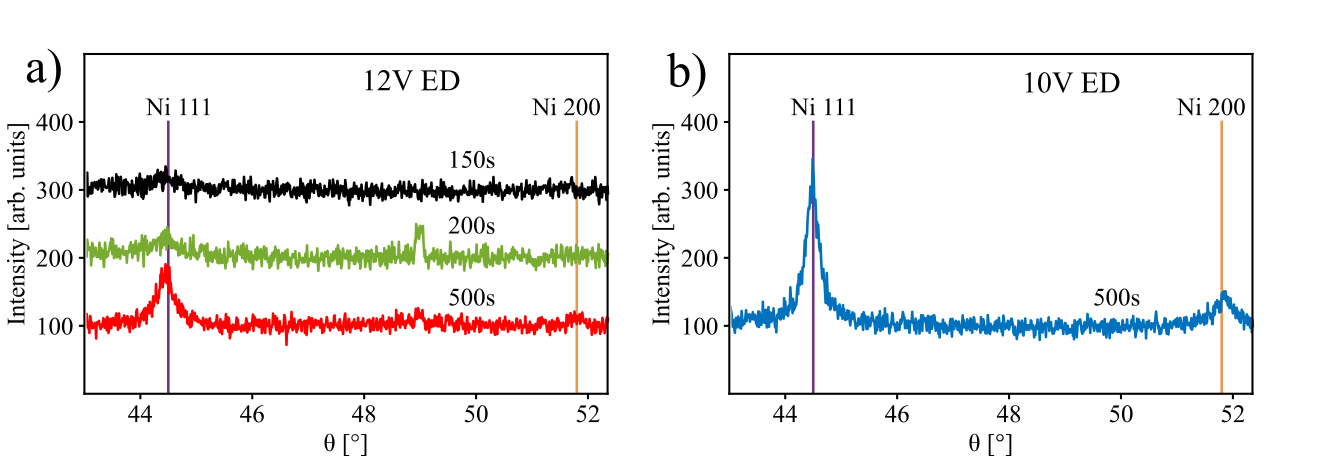}
 \caption{XRD patterns obtained from: a) 12 V etched samples and b) a 10 V etched sample. Electrodeposition times are noted next to the patterns. Ni (111) and (200) lines from ICSD 04-010-6148. Spectra have been shifted for clarity. We were unable to identify the source of the intermittent peak at $2\theta=49^{\circ}$.}\label{fig:XRD}
\end{figure}

\Fref{fig:XRD} shows a series of XRD patterns obtained from the samples. \Fref{fig:XRD}a shows the XRD evolution with electrodeposition duration for 12 V etched samples. The dominant crystallite orientation was (111) with a clear increase for longer depositions. The (200) peak was hardly visible even for the longest deposition, with a 10$\pm1$:100 integrated (200):(111) peak ratio found; a significantly lower ratio compared to powder diffraction (45:100). Comparing H500 and L500 samples (\Fref{fig:XRD}a,b), L500 shows a higher (111) peak, and a well defined (200) peak, with a ratio of 16$\pm1$:100. This indicates (111) preference in our electrodeposition conditions, and the maintenance of (111) preference for infiltrated growth in the porous samples. This result is in agreement with previous reports\cite{TANG2008,Boubatra2011morphology}, particularly when considering the 4.5 pH used here\cite{Boubatra2011morphology}.\\ \indent

By applying Scherrer's equation to the fitted peaks\cite{Le2008powder}, without particularly accounting for other sources of peak broadening, we obtained a lower estimate for the mean crystallite size. This was performed for the strongest signals and the results are shown in \Tref{tbl:XRD}. The results obtained for L500 (thin film structure), with a lower estimate of 40 nm, are in good agreement with previous reports on electrodeposited nickel crystallite size in this pH\cite{TANG2008,Boubatra2011morphology}. When comparing the infiltrated (12 V) and the thin film (10 V) electrodeposited structures, the latter exhibited larger crystallites, supporting our assumption that the porous structure could limit the growth.\\ \indent
We further examine the residual strain found in the nickel crystallites, based on a d-spacing of 2.037 \AA\ (or 1.764 \AA\ for (200), where appropriate; after ICSD 04-010-6148). Interestingly, nickel deposited on the 12 V samples results in a slightly higher strain compared to the 10 V samples (0.06\% compared to -0.08\%, correspondingly). However, both values are low compared to reported values in the literature (0.24\% on gold\cite{Boubatra2011morphology}) and are close to our confidence level in this estimate. The sample porosity strongly affects its mechanical properties\cite{Calahorra2020enhanced}; this could explain the different residual strain between samples. An overall compliant substrate can explain the reduced strain compared to previously reported values on non-porous substrates. We note that the exact values depend upon the choice of reference, however, even with a different reference used (2.034 \AA\cite{Dzhumaliev2012effect} or 2.03 \AA\cite{Sharma2016influence}) the trends described above persist, and the strains we found were small.\\ \indent
\begin{table*}[!htb]
\small
\center
  \caption{The crystalline properties obtained by XRD from samples with noticeable peaks. $L_{min}$ is a lower estimate for the Ni(111) crystallite size with a 5 nm error estimate, $\overline{S}$ is the measured strain percent, PR is the integrated (200):(111) peak ratio.}
  \label{tbl:XRD}
\begin{tabular}{>{\centering\arraybackslash}m{0.18\textwidth}>{\centering\arraybackslash}m{0.15\textwidth}>{\centering\arraybackslash}m{0.15\textwidth}>{\centering\arraybackslash}m{0.1\textwidth}}
 \toprule
    Sample &${L_{min}}$ [$\pm5$ nm]& $\overline{S}$ [$\pm0.05$\%] & PR\super{a}[$\pm1$\%] \\
    \mr
   {L500} &  $\sim${40} & -0.08 & 16\\ 
   {L500}\sub{(200)} & $\sim${40} & -0.14 & \\ 
   {H500} &   $\sim${30} & 0.06 & 10\\ 
 \bottomrule
 \end{tabular} \\ \flushleft
 \textit{a} - Powder (200):(111) peak ratio is 42.5\%.\\
\end{table*}
\subsection{Magnetization of nickel deposited on porous GaN}
\begin{figure}[th]
 \centering
 \includegraphics[scale=0.67]{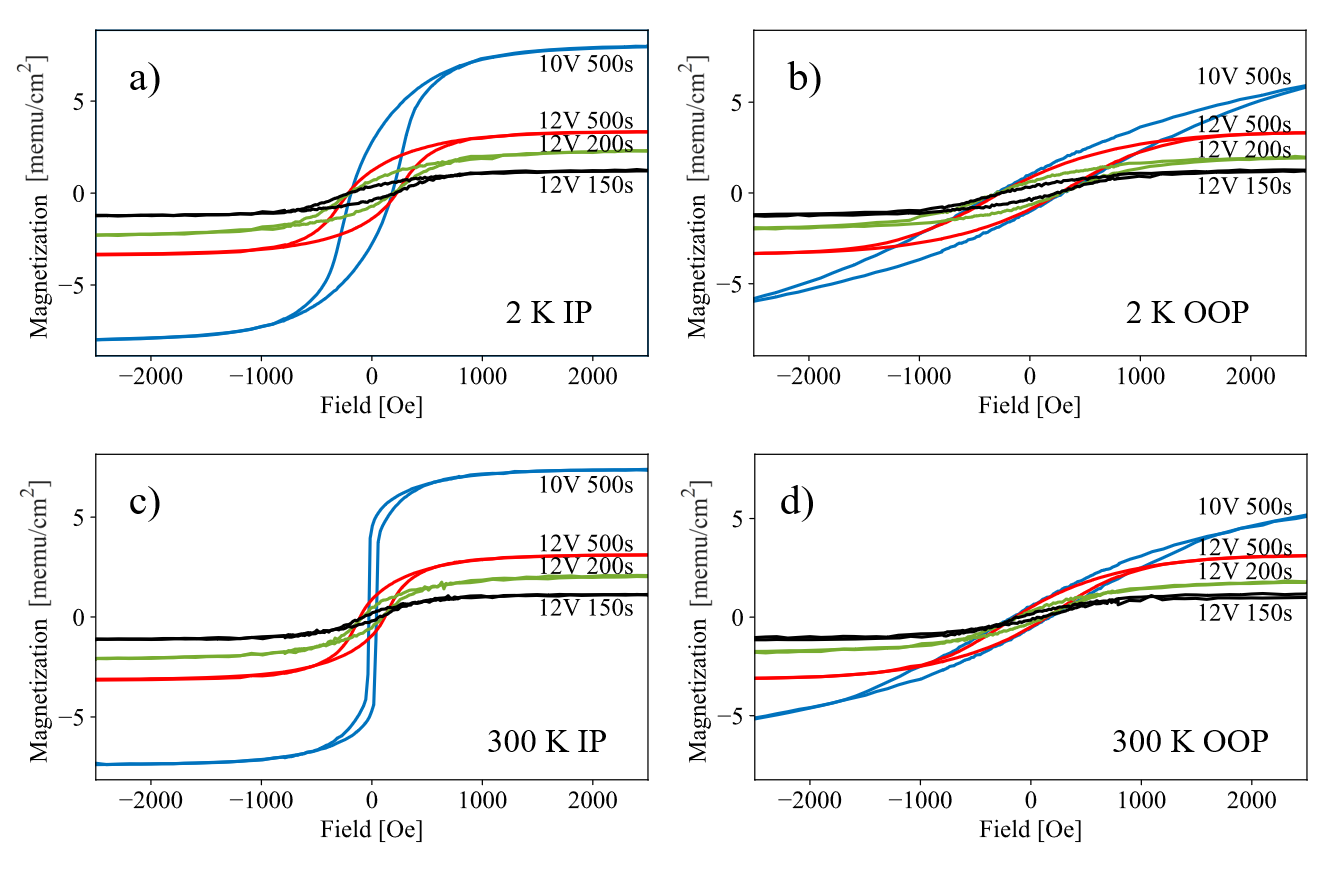}
 \caption{Lowest and highest temperature in-plane (IP) and out-of-plane (OOP) VSM results for L500, H500, H200, H150: a) IP magnetization at 2 K; b) OOP magnetization at 2 K; c) IP magnetization at 300 K; d) OOP magnetization at 300 K. Magnetization is normalized to sample area due to uncertainty in total nickel volume. For SI units: memu/cm\super{2} = 0.01A; 1000 Oe \tilda{} 80 kA/m.}\label{fig:VSM_low_field}
\end{figure}

The magnetization was characterized by a series of VSM measurements, taken in-plane and out-of-plane, relative to the GaN surface, at different temperatures. The temperatures examined were 2, 50, 100, 200 and 300 K, and the maximum external field was (at least) 4000 Oe. \Fref{fig:VSM_low_field} shows the (low field) in-plane and out-of-plane area-normalized magnetization loops recorded at 2 and 300 K. These curves show the main magnetization trends: the thin-film sample (L500) exhibited a pronounced easy/hard axis behaviour - as expected for a thin film. The easy axis was found in-plane, despite the dominant (111) crystalline orientation of the samples, indicating that shape anisotropy dominates the magnetization and not crystalline anisotropy. The in-plane easy-axis trend was weaker for 12 V etched samples, where infiltration of nickel was found. This was most prominent for samples of shorter electrodeposition times. These observations can be explained by the dispersed characteristic of the infiltrated nickel, forming nanoparticles, as seen in the SEM images. This can contribute to the diminished role of shape anisotropy of the overall structure
. See Supporting Information Figure S3 for the full corresponding magnetization loops recorded.\\ \indent

\begin{figure}[tb]
 \centering
 \includegraphics[scale=0.67]{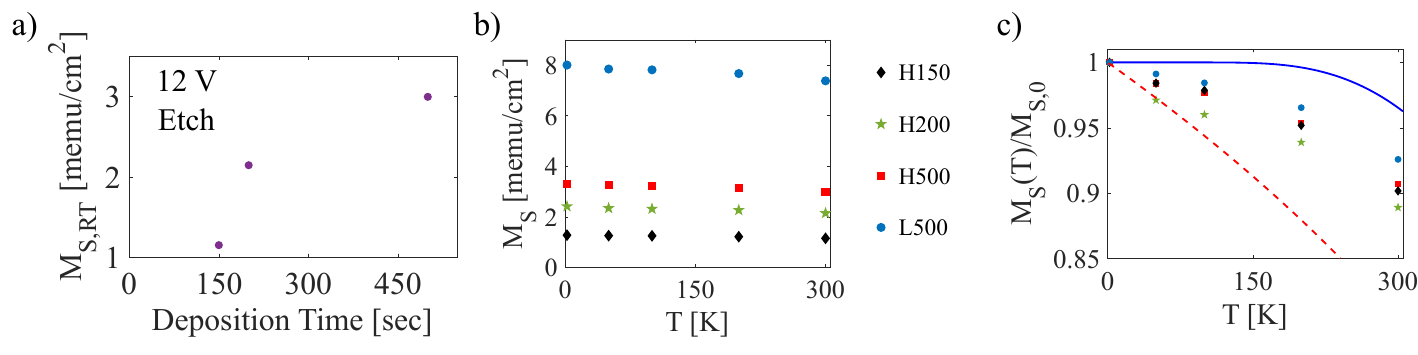}
 \caption{Saturation magnetization per unit area: a) for 12 V etched samples at 300 K as a function of deposition duration; b) as a function of temperature for all samples; c) relative to $M_{S,2K}$ and compared to theoretical Brillouin ($J=1/2$, blue solid) and Langevin (red dashed) curves with a Curie temperature of 628 K. [b,c] share the legend.}\label{fig:VSM_MST}
\end{figure}

\Fref{fig:VSM_MST}a shows the (area normalized) saturation magnetization ($M_S$) at 300 K as a function of electrodeposition duration for H150/200/500 samples. The saturation magnetization trend with time is nonlinear. The non-linearity in the electrodeposition saturation magnetization is congruent with the non-linear characteristics of the current profiles (\Fref{fig:ED_currents}b), which is indicative of the amount of nickel.\\ \indent
\Fref{fig:VSM_MST}b shows the temperature-dependent saturation magnetization per unit area of all the VSM measured samples. Sample L500 is unique since nickel was deposited as a layer with almost no infiltration. Therefore it can serve for quantifying the magnetization. Looking back at \Fref{fig:SEM}a, the layer thickness was measured as $155\pm38$ nm (based on several line measurements). Using 8.908 g/cm\super{3} for nickel density a saturation magnetization of $56.5\pm15$ emu/g (Am\super{2}/kg) was obtained for the 2 K series - in excellent agreement with expected value for nickel of about 58 emu/g\cite{Danan1968new}. The high quality of the electrodeposited nickel was also reflected in the XRD measurements as seen above.\\ \indent 
As mentioned above, we hypothesize that the conductivity of the remainder of the GaN layer had a significant role to play in determining the characteristics of the deposited nickel - resulting in a top layer for smaller pores/higher conductivity sample (L500) and infiltrated structures for larger pores/lower conductivity samples. The lower saturation magnetization per cm\super2 obtained for H500, supports this hypothesis (lower conductivity corresponding to less material deposited). This is under the assumption that the electrodeposited nickel has similar properties regardless of shape.\\ \indent
\Fref{fig:VSM_MST}c shows the saturation magnetization normalized to the 2 K value for electrodeposited and sputtered samples, correspondingly. For reference, the Brillouin (with quantum number J=1/2) and Langevin functions with a Curie temperature of 628 K are also shown. The magnetization evolution with temperature was found to lie between the quantum and classical limits, with the data approaching the Langevin function for shorter/thinner depositions. This could be an indication of a transition from thin film to a more isotropic nano-particle magnetization regime\cite{Spaldin2010}.\\ \indent

\begin{figure}[th]
 \centering
 \includegraphics[scale=1]{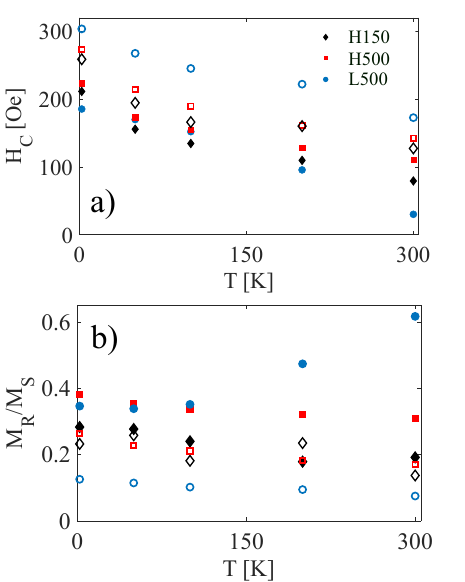}
 \caption{Magnetization properties of electrodeposited samples: a) Coercivity; b) remanence. The legend is relevant to both figures, closed/open symbols represent IP/OOP loops.}\label{fig:VSM_HCMR}
\end{figure}

\Fref{fig:VSM_HCMR} shows the in-plane and out-of-plane coercivity, $H_C$, and remanence, $M_R$, of the electrodeposited samples for the entire range of temperatures measured. Samples H200 and H150 displayed very similar coercivity and we present here only sample H150 to improve figure clarity. Sample L500 demonstrated in-plane easy axis characteristics, with lower in-plane coercivity and higher remanence. The XRD patterns revealed a dominance of (111) oriented growth - which is the magnetocrystalline easy axis for fcc crystals\cite{Spaldin2010}. It also revealed relatively low strains in all samples, indicating low magnetoeleastic contribution, and we can conclude that the thin film shape anisotropy dominates the magnetization of sample L500. We compare the properties of L500 to similar thin films grown by Boubatra and co-workers\cite{Boubatra2011morphology}. They found higher coercivity values in their films (120-150 Oe compared to 30 Oe here at room temperature), despite growth in similar conditions, exhibiting similar crystallinity. They attribute the high coercivity to strain in the grown films. This interpretation is in agreement with the low strains we found in our films, probably due to relaxation enabled by the porous substrate. The in-plane remanence of L500 increased with temperature to a value of 0.6 at 300 K. The complete reasons for that are unclear and require further studies.\\ \indent
The magnetization of infiltrated samples was different compared to the thin film. A clear trend emerged when comparing samples H150/500 and L500. Firstly, the infiltrated samples demonstrated diminished anisotropy. This is manifested in the reduction of the in-plane/out-of-plane difference in coercivity and remanence. Furthermore, the out-of-plane saturation field of the infiltrated samples was considerably lower (Supporting Information Figure S3.1). Overall, the results obtained for both 12 V samples (H150/500) are in between the in-plane/out-of-plane results of L500. This trend intensifies in sample H150 where the in-plane and out-of-plane remanence were the closest of all samples measured (still with slight in-plane preference), and the coercivity of H150 is lower than that of H500. In the infiltrated structure several factors are at play, including particle size and packing. The lower coercivity of H150 can indicate a reduction of particle shape anisotropy. It can also indicate that following initial nucleation, packing density did not significantly increase with time and that the particles preferentially grew, rather then nucleate (since increased density is associated with reduced coercivity\cite{Spaldin2010}). Furthermore, as seen in SEM (\Fref{fig:SEM}) and below in MFM (\Fref{fig:MFM}) sample H500 is partially covered by a thin film and its mixed structure should indeed show in magnetization properties. These results open the possibility of engineering effectively isotropic magnetic thin films layers by controlling the etching and deposition within porous media.\\ \indent

\subsection{MFM studies}
\begin{figure}[tb]
 \centering
 \includegraphics[scale=0.6]{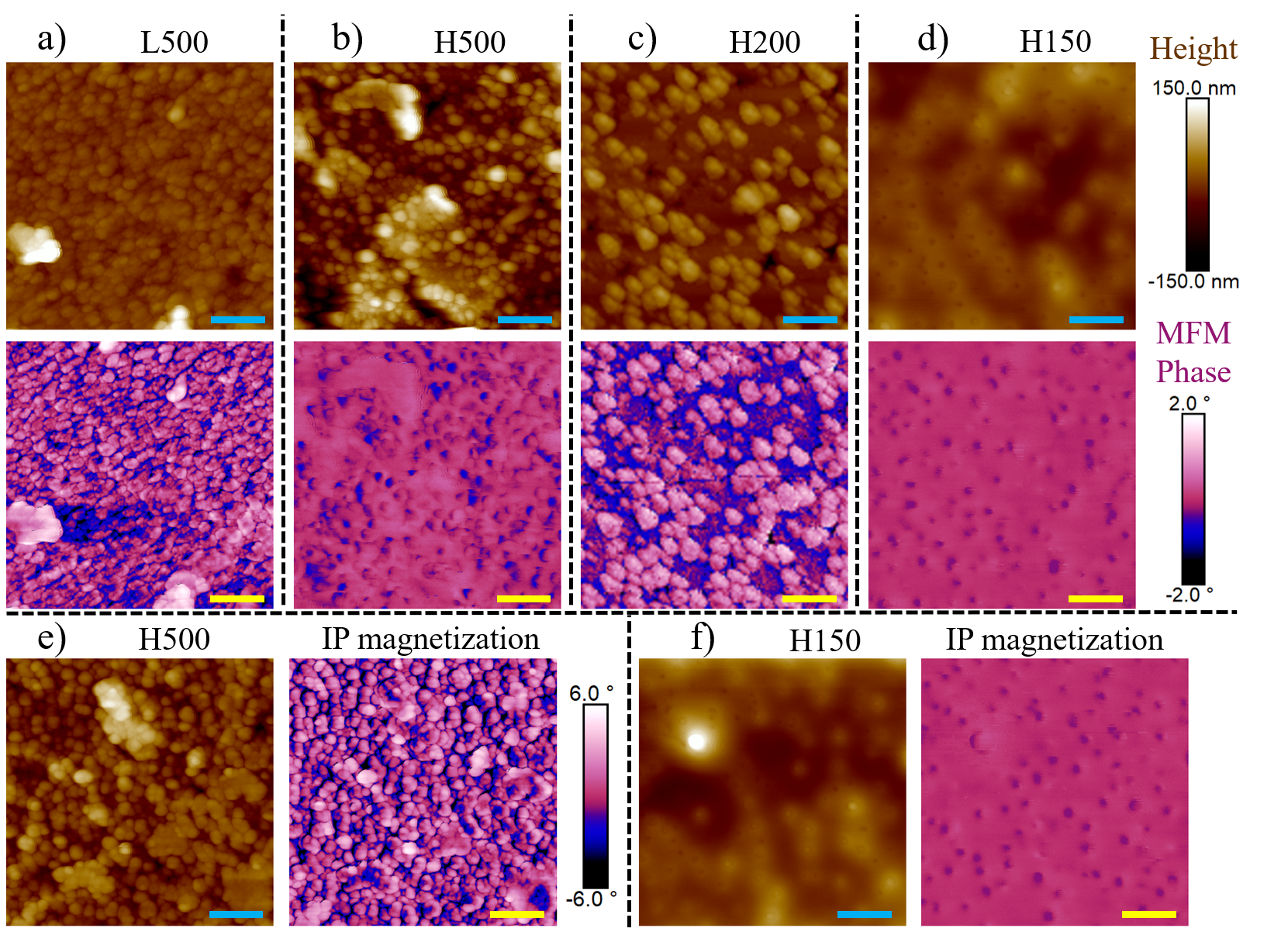}
 \caption{AFM topography and MFM phase of the studied samples: a-d) topography (top) and MFM phase (bottom) recorded following OOP magnetization of the samples; e,f) topography (left) and MFM phase (right) of samples H500 and H150, recorded after IP magnetization. All scale bars are 800 nm, color scales shown on the right are relevant to all panels except for MFM [e]. Sample H200 was measured using PF-MFM.}\label{fig:MFM}
\end{figure}

\Fref{fig:MFM} shows the topography and corresponding MFM phase signals of the studied samples. The topography was found to be very well correlated to the SEM images. In particular, samples H500/200/150 show the evolution of the surface with deposition time: from a smooth surface with visible pores (H150) to the protrusions found after longer duration deposition (H500). The root-mean-square roughness values calculated for these samples were: 12.5 nm for L500 (excluding the bright contamination, 24 nm including it), 44 nm for H500, 24 nm for H200 and 19 nm for H150. These results further corroborate our understanding of the electrodeposition processes depending on pore size and duration.\\ \indent
The MFM phase images show the different magnetic properties of the samples. When comparing samples H500 and L500 after OOP magnetization, distinct patterns were measured despite both samples were covered by a layer of nickel. Interestingly, the L500 grain pattern, as reflected in topography, exhibited good correlation with the OOP magnetization MFM phase, while the H500 did not show this correlation, even though the grains/particles formed a rougher topography. Sample H200 exhibited a surface comprised of nickel protruding from inside the pores and bare GaN surface; the MFM image shows the magnetized nickel particles. The surface of sample H150 had pores and not protrusions, indicating that any nickel was indeed infiltrated and did not contribute to the MFM signal.\\ \indent
Panels [e] and [f] show topography and MFM phase of samples H500 and H150 after in-plane poling. Importantly, the measurement itself was carried out in the same manner, \textit{i.e.}, the vertical signal of AFM vibration (more sensitive to OOP interactions) was detected. The distinction between IP and OOP magnetization in H500 is striking (panels [b,e]): the former shows the particles tend to magnetize individually, and the magnetization direction (horizontally in this case) shows a strong contrast within the domain/particle. Notice that the phase difference in this case was considerably larger. This observation is in agreement with the higher remenance of the IP response in this sample (\Fref{fig:VSM_HCMR}). It is also interesting to compare the MFM phase signals of sample H150 after OOP and IP magnetization (panels [d,f]). In this case no distinction was observed (both exhibited similar MFM phase images), in agreement with the sample's infiltrated structure and the lowest observed easy/hard axes anisotropy in VSM (\Fref{fig:VSM_low_field}): no top thin-film means no contribution from this kind of structure to the IP/OOP anisotropy.\\ \indent

\section{Summary}
We studied the structural and magnetic properties of Ni/porous-GaN composites prepared by electrodeposition onto/into the porous GaN layer. Our results indicate that the magnetic response is intricately dependent on the amount of nickel deposited and the pore size. In particular, we found that electrodeposition on porous materials is characterised by low strains and dominant (111) out of plane orientation of the deposited nickel. Electrodeposition was found to lead to good crystallinity even without subsequent treatments and allowed for realization of nickel-infiltrated composites. These nickel-infiltrated composites exhibited low magnetic anisotropy despite their overall thin-film form. Deposition on a sample of originally smaller diameter pores resulted in a thin film with pronounced easy/hard axis characteristics, as expected for thin-films, and low coercivity - which we attribute to low stresses in the film. These results direct further investigations on the ability to control the magnetic properties of nanostructures and of magnetic/semiconducting material composites to develop new applications coupling these properties.
\section*{Data and Supporting Information}
The data that support the findings of this study are openly available at http://  
Supporting information is available on the publisher website.  

\section*{Acknowledgments}

\section*{References}
\bibliography{current.bib}
\bibliographystyle{unsrt}

\newpage
\title[]{Magnetic properties of nanoporous GaN and Nickel composites prepared by electrodeposition}

\section*{\Large Supplementary Information}
\clearpage
\paragraph*{Section S1. GaN growth and Etching\\}
Figure S1 shows cross-sectional SEM images of  about 1 $\mu$m porous GaN etched at different voltages, of differently doped samples. Etching is stronger for higher doping and higher voltages, and the resulting effect is visible to the naked eye, in light reflectance properties.
\begin{center}
     \includegraphics[scale=1.4]{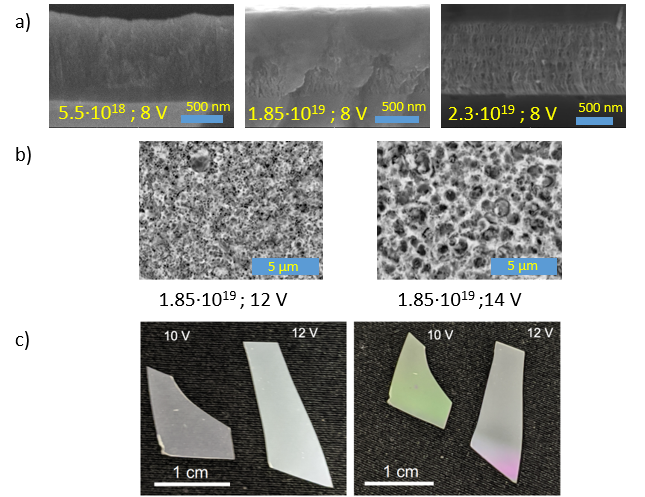}\\
\end{center}
Figure S1. a) Cross-sectional SEM image showing that porosity increased with voltage; b) top-view SEM image showing increased etching with voltage; c) optical images of samples similar to the ones reported, doped 1.85$\times10^{19}$ cm$^{-3}$ and etched at 10 and 12 V. The samples reflect light differently, following the different etching. These images show the same samples from different angles. 

\clearpage
\paragraph*{Section S2. Nickel electrodeposition\\}
Calibration depositions were made using  ITO coated PET. The relation between thickness and deposition voltage was found to be highly non-linear, while it was linear with deposition time. This also reflected in magnetization, where the magnetization vs. the total charge passed in the process also showed a linear trend.
\begin{center}
     \includegraphics[scale=0.7]{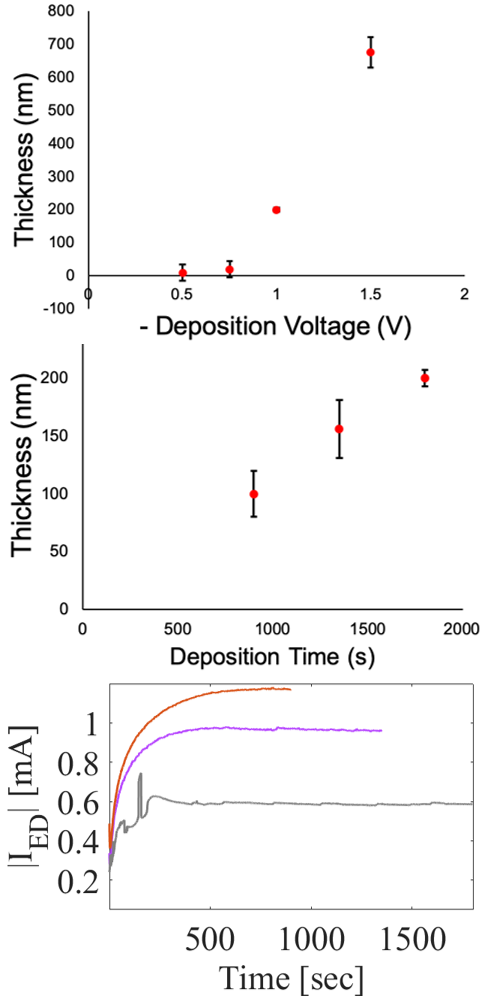}\\
\end{center}
Figure S2.1 Thickness of electrodeposited nickel as vs. voltage (top, for 1800 s), and vs. time (middle, using -1 V). The bottom image shows control current evolution curves. 
\begin{center}
 \includegraphics[scale=0.7]{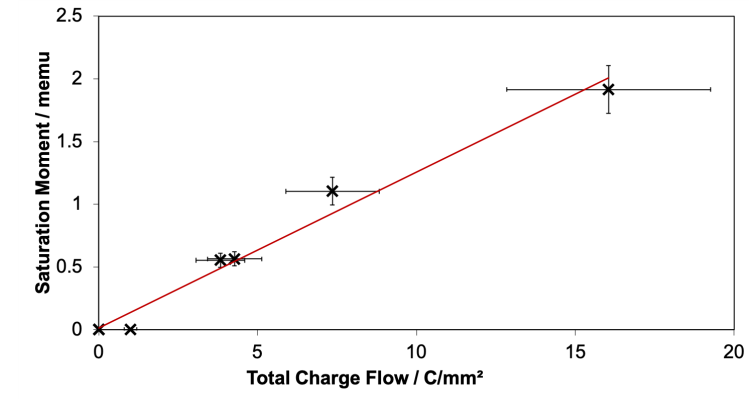}
 \end{center}
Figure S2.2 Magnetization vs. electrodeposition charge obtained by integrating over the current. 

\begin{figure}[h]
 \begin{center}
 \includegraphics[scale=0.45]{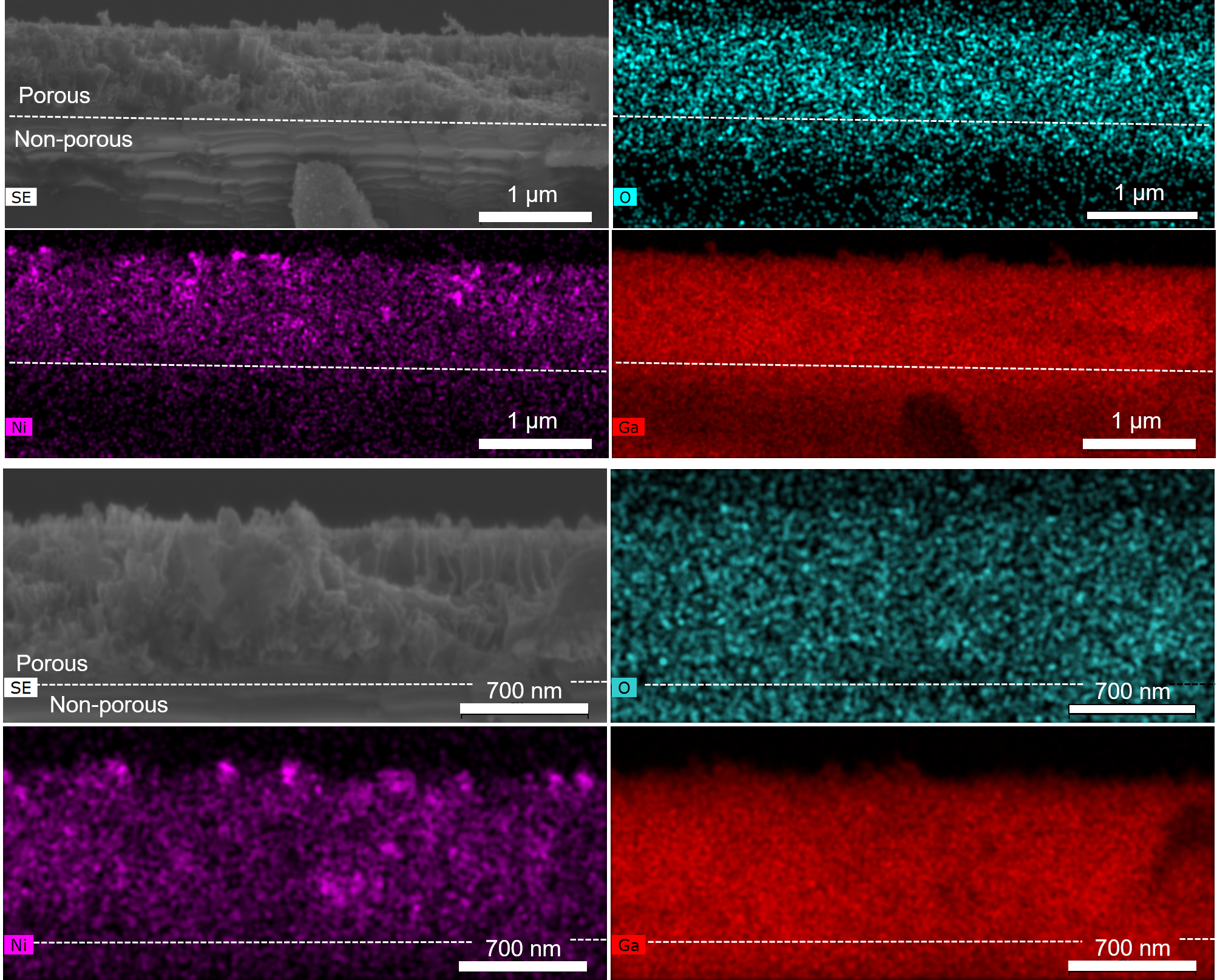}
 \end{center}
 {Figure S2.3 SEM-EDX data obtained from two different locations of H100. No significant distinctions along the sample cross section were found.}
\end{figure}
\clearpage
\paragraph*{Section S3. Magnetization loops\\}
Figure S3 shows the full range magnetization loops measured at 2 K and 300 K in-plane and out-of-plane.
\begin{center}
\includegraphics[scale=0.6]{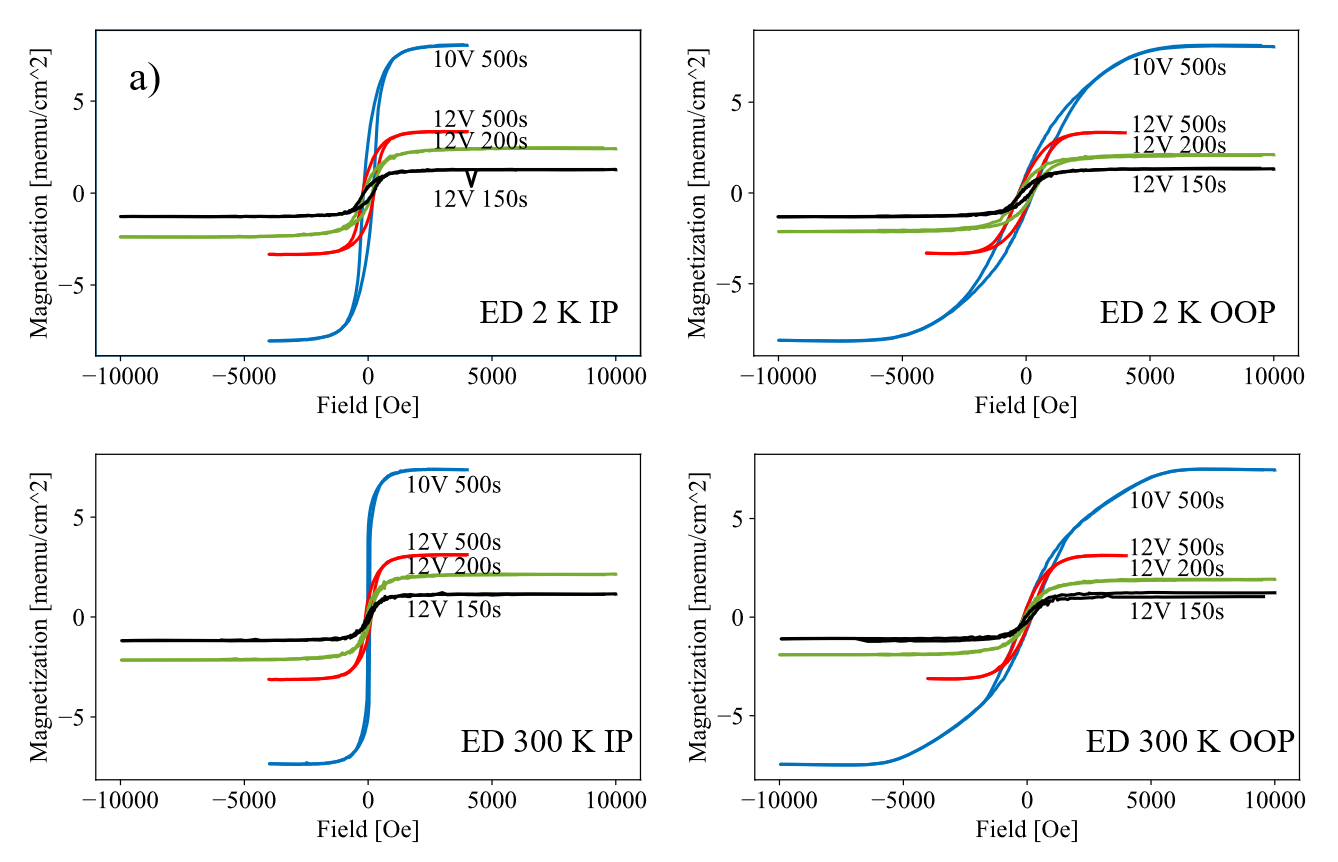}
\end{center}
\clearpage

\end{document}